\documentclass[twoside]{article}

\usepackage{PRIMEarxiv}

\usepackage[utf8]{inputenc} % allow utf-8 input
\usepackage[T1]{fontenc}    % use 8-bit T1 fonts
\usepackage{hyperref}       % hyperlinks
\usepackage{url}            % simple URL typesetting
\usepackage{booktabs}       % professional-quality tables
\usepackage{amsfonts}       % blackboard math symbols
\usepackage{nicefrac}       % compact symbols for 1/2, etc.
\usepackage{microtype}      % microtypography
\usepackage{lipsum}
\usepackage{fancyhdr}       % header
\usepackage{graphicx}       % graphics
\usepackage{tabularx} % Import the tabularx package
\usepackage{multirow}
\usepackage{subfigure}
\usepackage[colorinlistoftodos]{todonotes}
\usepackage{amsmath,amssymb}

\NewDocumentCommand{\halfcirc}{ O{} }{%
    \begin{tikzpicture}
        \fill[white] (0,0) circle (1.0ex); 
        \fill[black] (0,0) -- (270:1ex) arc (270:90:1ex) -- cycle; 
        \draw[black] (0,0) circle (1.0ex);
    \end{tikzpicture}
}
 
\NewDocumentCommand{\rot}{O{60} O{1em} m}{\makebox[#2][l]{\rotatebox{#1}{#3}}}%
\usepackage{pifont}% http://ctan.org/pkg/pifont
\newcommand{\cmark}{\ding{51}}%

\title{Service Level Agreements and Security SLA: A Comprehensive Survey}

\author{
  Serena Nicolazzo\\
  Department of Computer Science,\\
  University of Milan, Italy \\
  \texttt{serena.nicolazzo@unimi.it} \\
  \And
  Antonino Nocera \\
  Department of Electrical, Computer and\\ Biomedical Engineering,\\ University of Pavia, Italy \\
  \texttt{antonino.nocera@unipv.it} \\
  \And
  Witold Pedrycz\\
  Department of Electrical and Computer Engineering,\\
  University of Alberta, Edmonton, AB T6G 2R3, Canada and\\
  Systems Research Institute, Polish Academy of Sciences
  00-901 Warsaw, Poland and\\
  Research Center of Performance and Productivity Analysis, Istinye University Istanbul, T{\"u}rkiye\\
  \texttt{wpedrycz@ualberta.ca}
}

\begin{document}
\maketitle

\begin{abstract}
A Service Level Agreement (SLA) is a formal
contract between a service provider and a consumer, representing a crucial instrument to define, manage, and maintain relationships between these two parties. The SLA's ability to define the Quality of Service (QoS) expectations, standards, and accountability helps to deliver high-quality services and increase client confidence in disparate application domains, such as Cloud computing and the Internet of Things. An open research direction in this context is related 
to the possible integration of new metrics to address the security and privacy aspects of services, thus providing protection of sensitive information, mitigating risks, and building trust. 
This survey paper identifies state of the art covering concepts, approaches, and open problems of SLA management with a distinctive and original focus on the recent development of Security SLA (SecSLA). It contributes by carrying out a comprehensive review and covering the gap between the analyses proposed in existing surveys and the most recent literature on this topic, spanning from 2017 to 2023.
Moreover, it proposes a novel classification criterium to organize the analysis based on SLA life cycle phases. This original point of view can help both academics and industrial practitioners to understand and properly locate existing contributions in the advancement of the different aspects of SLA technology.
The present work highlights the importance of the covered topics and the need for new research improvements to tackle present and demanding challenges.
\end{abstract}

% keywords can be removed

\section{Introduction}

Establishing clear and shared rules between consumers and providers to regulate IT service provisioning is a key factor for guaranteeing the expected quality of service and increasing parties' protection in case of disputes or disagreements. 
This is even more true in modern Cloud computing and Internet of Things (IoT) environments characterized by high dynamicity and the intrinsic heterogeneity of their components.

To tackle this problem, Service Level Agreements (SLAs, for short) have been proposed as an effective method to manage and guarantee the promised Quality of Service (QoS), accurate reporting on service usage, and runtime adaptation for evolving requirements. 

An SLA is a formal and legally binding contract or agreement defining the terms and performance metrics that the service provider commits to delivering and the customer expects to receive.
Therefore, from the providers' side, an SLA ensures that they can avoid penalties if the appropriate levels of agreement have been respected during the service provisioning, and, at the same time, providers can strengthen their credibility if the given conditions have been always met. On the other hand, also customers can benefit from these contracts because they can be assured that the favored service will follow the chosen terms \cite{kyriazis2013cloud}.

Especially in the above-cited Cloud and IoT contexts, SLA management gives rise to several challenges. These can be related to 
the lack of transparency on the actual performance of the services (which can vary also for external conditions), little control over infrastructures, multi-tenancy heterogeneous hardware, lack of standardization and specific regulatory requirements from different industries and regions, and, finally, the increasing number of connected devices that gives rise to scalability problems\cite{girs2020systematic,noureddine2021ml}.
Still in this context, a major challenge is related to the vulnerability to security and privacy threats, which may lead to information disclosure and service
unavailability. On the other hand, attacks on infrastructures and devices directly impact the goals of SLAs \cite{bhonde2021impact}. While security has long been considered a potential element of SLAs \cite{henning1999security}, establishing measurable and commonly agreed-upon security metrics has always been considered a hard task for researchers and practitioners. Consequently, identifying approaches to effectively manage security throughout all the phases of the SLA life cycle requires a systematic review of existing contributions and additional research efforts in this direction \cite{casola2015adoption}.

The main goal of this survey paper is to carry out a detailed and critical investigation of the existing research on the management of SLAs.
To do so, we retrieve
relevant studies focusing on various aspects of SLA management and propose a classification based on the SLA life cycle discussed in the European Commission Report \cite{kyriazis2013cloud}. Starting from this, by analyzing the contributions available in the literature, we can identify three main categories, covering SLA modeling, SLA negotiation, and SLA monitoring and violation, which we can use to organize our systematic analysis. As an additional important contribution, in our proposal, we focus on approaches leveraging Security Service Level Agreements (SecSLA, hereafter), which attempt to integrate security metrics into SLAs. Moreover, still, in this context, we consider Privacy Level Agreements (PLAs), which are a subcategory of SecSLAs specifically dealing with notions related to privacy protection practices.
Then, we evaluate the existing literature maturity level and find commonalities and research gaps also drawing possible future directions and open issues for the research community.

The contribution of this study can be classified into the following four main folds:
\begin{itemize}
    \item we identify the most prominent and recent studies related to SLA and Security SLA.
    \item We provide a comprehensive analysis of the main basic concepts related to SLA, such as SLA definition, life cycle, and language specifications.
    \item We undertake a complete examination of the identified approaches and organize our analysis based on the SLA life cycle phases.
    \item We thoroughly examine open issues, challenges, and future trends of SLA and SecSLA approaches.
\end{itemize}

We believe that this paper can contribute to the academic discourse and industrial progress in identifying the open challenges in this context and providing a clear and updated picture of the SLA and SecSLA landscape.
Although SLA and SecSLA play a crucial role in defining shared rules and security metrics for services at design time,
nevertheless, a comprehensive survey on recent advancements in SLA and SecSLA is missing in the current scientific literature. 
Moreover, by organizing our analysis using the SLA life cycle phases to group related proposals, this survey paper can help academics and industrial practitioners to understand and properly locate the contributions, available in the recent literature, to the different aspects of the SLA technology.

The remainder of this paper is organized as follows.
Section \ref{sec:methodology} focuses on the methodology employed to conduct this work, whereas Section \ref{sec:related} discusses the articles related to our survey. In Section \ref{sec:background}, we overview the main concepts related to Service Level Agreements, namely SLA definition, characteristics, actors, life cycle, and classical language specifications. Section \ref{sec:SLA} reviews the literature relying on our proposed classification method focusing on the SLA life cycle aspects of modeling, negotiation, monitoring, and violation.
Section \ref{sec:secSLA} examines the existing scientific contributions from a security and privacy perspective describing works dealing with SecSLAs. Section \ref{sec:future} delves into the open problems and future research directions. Finally, Section \ref{sec:conclusion} provides a summary of the present survey, offers some perspective and opportunities, and concludes this work.

\section{Methodology}
\label{sec:methodology}
\begin{figure}
    \centering
    \includegraphics[width=0.8\linewidth]{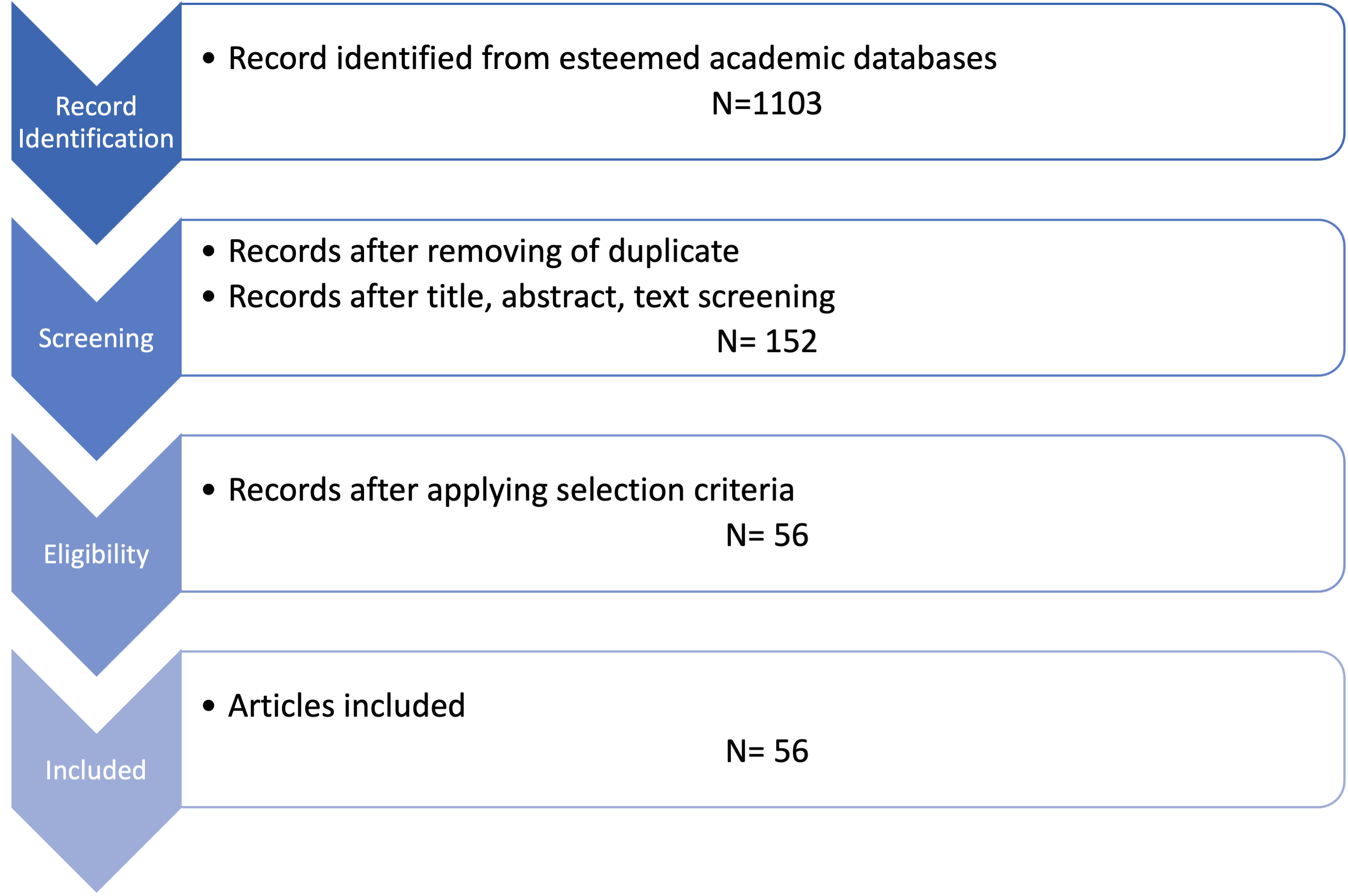}
    \caption{PRISMA flowchart for paper selection process}
    \label{fig:prisma}
\end{figure}

In this paper, we conduct a survey study following the guidelines drawn in \cite{petersen2015guidelines} to identify published research results that are relevant to the research area of interest. 
We accomplish a comprehensive literature review process, following these steps: {\em(i)} defining clear search objectives, {\em(ii)} identifying relevant literature from journals and conference proceedings leveraging search engines and driven by a well-defined search strategy, and {\em(iii)} applying selection criteria to filter the articles during this iterative process. Figure \ref{fig:prisma} shows the PRISMA flow diagram, providing a visual representation of the above-described screening process. The diagram highlights the count of research works identified, excluded, and included.

\subsection{Search strategy}

To gather relevant research focusing on SLA, we designed a search strategy to align with our research objectives. We meticulously performed searches through esteemed academic databases, such as {\em (i)} IEEE Xplore Digital Library \cite{ieeeXplore}, {\em (ii)} Scopus \cite{Scopus}, {\em (iii)} Google Scholar \cite{Scholar}, and {\em (iv)} ACM Digital Library \cite{ACMDigLib}. The search scope comprises a broad range of publication years from 2017 to 2023, ensuring a well-rounded coverage of recent research in the field. Furthermore, we developed the search terms to identify relevant investigations and employ a combination of specific keywords and phrases to encompass various aspects of SLA. The primary search terms include ``Service Level Agreement", and ``SLA". In conjunction with the primary terms we used additional keywords such as ``SecSla", ``PLA", ``SLA management", ``IoT", ``Web Service'', and ``Cloud Computing". 

Starting from these terms, we formulated the following search string:

\centerline{(``Service Level Agreement'' OR ``SLA'' OR ``SLA Management'')}
\centerline{AND} 
\centerline{(``Security Service Level Agreement'' OR ``SecSLA'' OR ``Security SLA'' OR}
\centerline{``Privacy Service Level Agreement'' OR ``PLA'')}
\centerline{AND} 
\centerline{(``Internet of Things’’ OR ``IoT'' OR ``Industrial
Internet of Things'' OR}
\centerline{``Cloud'' OR ``Cloud Computing'' OR ``Web Service'')}

\subsection{Selection criteria}
In this section, we list the selection criteria we used to decide whether a scientific work, identified through the above-described search query, is appropriate for our study and reaches enough quality to be included in this survey. A paper is eligible for inclusion in the present work if it satisfies at least one of the inclusion criteria and none of the exclusion criteria applies.
At the end of this screening, 56 papers were finally selected.

\subsubsection{Inclusion Criteria}

To evaluate the relevance of a paper and include it in our survey, we observed the following criteria:

\begin{itemize}
    \item the corresponding author or the supervisor's importance in the field under analysis;
    \item the importance of the venue (journal or conference) where the paper has been published (we consider Scimago \cite{Scimago} and Core.edu \cite{Core} as ranking Web sites for journals and conferences, respectively). 
\end{itemize}

\subsubsection{Exclusion Criteria}

After the inclusion, the exclusion process is followed. A paper is excluded if one of the following criteria is met:

\begin{itemize}
    \item the paper is not written in English;
    \item the paper is not peer-reviewed;
    \item the paper's year of publication is earlier than 2017;
    \item the paper is not specifically focused on SLA nor contributes to this context;
    \item the paper lacks relevance, it is an incremental refinement of an earlier proposed approach, there is a duplicate publication, or there is a more recent and cited version of the work published in a prestigious venue.
\end{itemize}

\subsubsection{Selection Process}
We used a two-step selection process to identify relevant
publications in our study as shown in Figure \ref{fig:prisma}. In the first step, we initially collected
$1,103$ possibly relevant publications, specifically: $290$ from IEEE Xplore Digital Library, $445$ from Scopus, $124$ from Google Scholar, and $264$ from ACM Digital Library.
Starting from this set, we deleted duplicates and performed a first screening of the abstract and text of the articles. 
Moreover, we also analyzed the bibliographies of the selected articles to identify possible additional interesting contributions.
This step led to $152$ publications left. Next, we applied the inclusion and exclusion
criteria defined before to classify papers relevant to our survey and not relevant ones. 
The relevant
publications were sorted according to the following categories: SLA Definition, SLA Modeling, SLA Negotiation, SLA Monitoring and Violation, and Security and Privacy SLA (see Section \ref{subsec:lifecycle}). This last step in the selection process shortlisted $56$ publications out of $152$ identified studies.
Therefore, this survey's final list of studies consists of $56$ publications.

\section{Related Work}
\label{sec:related}

In this section, we provide an overview of the existing research literature in the field of SLA, SLA management, and Security SLA. 

The survey paper presented in \cite{girs2020systematic} mainly focuses on the definition and SLA modeling for Cloud services in IoT. They analyze 44 publications between 2017 and 2018. The classification scheme used to group the papers refers to the following groups: {\em (i)} ontology, {\em (ii)} languages, {\em (iii)} frameworks and methodologies, {\em (v)} templates and {\em (vi)} matching of services and consumer needs.
The papers \cite{mubeen2017management,papadopoulos2017slas} deal with the management of SLAs in IoT applications. In particular, \cite{mubeen2017management} identifies
$328$ studies and categorizes them into seven technical classifications of the SLA life cycle. Instead, the work of Papadopoulos et al. \cite{papadopoulos2017slas} conducts a systematic mapping study to allow for a concise understanding of the status of SLA management in industrial IoT analyzing works between 2012 and 2016.

Several works focus only on a specific aspect of the SLA life cycle \cite{he2018review,prasad2018efficient,hamdi2021survey}.
For instance, the survey article in \cite{he2018review} focuses particularly on the SLA negotiation phase in Cloud computing studying a few research contributions from 2009 to 2017.
Instead, the work \cite{prasad2018efficient} examines 23 papers related to monitoring and prediction techniques for the efficient usage of IaaS Cloud resources.
A recent contribution \cite{hamdi2021survey} focuses on approaches based on Blockchain technology. This survey analyzes papers, published over the period from 2018 to 2020, which employ this technology to provide trust between consumers and service providers based on SLA violations and compensation enforcements.

The objective of the surveys proposed in \cite{odun2017cloud,odun2019cloud}
is to examine Cloud service computing and SLA. In particular, \cite{odun2017cloud} discusses briefly Cloud SLAs, their issues, and developmental challenges providing also a description of the major commercial Cloud providers. 
The authors of \cite{odun2019cloud}, instead, document techniques for scheduling and resource allocation in the Cloud.

A recent survey paper from Sharma et al. \cite{sharma2023sla} deals with intent-driven service management (IDSM) systems. In this context, intent is defined as a declarative expression representing what a user wants to achieve instead of how it should be achieved. The authors focus on works related to SLA management, specifying SLAs as intents. Moreover, a taxonomy is proposed and used to compare the analyzed techniques in IDSM systems.

A few works analyze SecSLA \cite{nugraha2017understanding,de2017state} with a focus on the open problems and existing solutions when integrating security properties into SLA contexts and, hence, providing a review of the literature on trustworthy SLAs. 

In comparison with the existing survey articles, analyzed in this section, our paper presents a complete discussion of publications produced in a very recent period (i.e., from 2018 to 2023) of all the phases of SLA Management (i.e., definition, modeling, negotiation, monitoring, and violation). Moreover, unlike earlier endeavors, our work delves into the security and privacy aspects of SLA, providing an investigation of recent analyses dealing with this perspective. 
This distinctive focus characterizes our paper concerning the existing body of literature, highlighting its originality and potential impact on the advancement of knowledge in the field.

Table \ref{tab:relatedSurveys} provides a summary analysis of the contributions of the existing related survey papers compared to ours.
In particular, this table illustrates the period of the analyzed publication, its type (i.e., survey paper or systematic review), the different phases of the SLA life cycle considered in the paper, the focus on security or privacy SLAs, and the domain of application.
Observe that, we refer to systematic reviews if the work provides only a synthesis of the analyzed results about the SLA topic, privileging numerics but not descriptions.

\begin{table}
\centering
\caption{Survey papers related to our work}
\footnotesize
\centering
\begin{tabular}{|l|lp{1.5cm}lllllp{2.6cm}|}
    \hline
    Paper & \rot{Literature timeline}& \rot{Paper Type}& \rot{SLA Definition}  &\rot{SLA Modeling} &\rot{SLA Negotiation}&\rot{SLA Monitoring and Violation} &
    \rot{SecSLA and PLA}&\rot{Domain} \\
    \hline \hline
    He and Sun \cite{he2018review} & 2009-2017 & Survey & - & - & \cmark& - & -& Cloud\\
    Papadopoulos et al. \cite{papadopoulos2017slas} & 2012- 2016 &
    Systematic Review & \cmark & \cmark & \cmark & \cmark & - & Industrial IoT\\
    Girs et al. \cite{girs2020systematic}& 2017-2018 & Survey &  \cmark & \cmark & - & - & - & Cloud-based IoT\\
    Mubeen et al. \cite{mubeen2017management}& 2009-2016 &  Survey& \cmark & \cmark & \cmark & \cmark & - & Cloud-based IoT\\
    Odun-Ayo et al. \cite{odun2017cloud}& 2009-2016 &  Survey & \cmark & - & - & - & - & Cloud\\
    Odun-Ayo et al. \cite{odun2019cloud}& 2008-2018 &  Survey & \cmark & - & - & - & - & Cloud\\
    Sharma et al. \cite{sharma2023sla}& 2016-2022 & Survey & \cmark & - & - & - & - & IDSM\\
    Nugraha and Martin \cite{nugraha2017understanding} & 1999-2017& Survey & - & - & - & - & \cmark & -\\
    De Carvalho \cite{de2017state} & 2009-2015 & Systematic Review & - & - & - & - & \cmark & Cloud\\
    Prasad and Bhavsar \cite{prasad2018efficient} & 2000-2017 & Survey & - & - & - & \cmark & - & Iaas Cloud\\
    Hamdi et al. \cite{hamdi2021survey} & 2018–2020 & Survey & - & - & - & \cmark & - & -\\
    \hline
    \textbf{Our Survey} & 2017-2023 & Survey & \cmark & \cmark & \cmark & \cmark &\cmark &IoT, Cloud, Cloud-based IoT, Web Services \\
    \hline\hline
\end{tabular}
\label{tab:relatedSurveys}
\end{table}

\section{Background}
\label{sec:background}

In this section, we provide the necessary background for this survey paper. In particular, we define what is an SLA and describe its main components, all the phases related to its life cycle, the defined language specifications, and the categorization employed throughout this work. 
Table \ref{tab:SystemSymbols} summarizes the acronyms used in this paper.

\begin{table}
\centering
  \caption{Summary of the acronyms used in the paper.\label{tab:SystemSymbols}}
  \begin{tabular}{ll}
    \toprule
    Symbol&Description\\
    \midrule
    IDSM & Intent-Driven Service Management\\
    IMS & IP Multimedia Subsystem\\
    IoT & Internet of Thing\\
    ML & Machine Learning\\
    PLA & Privacy Service Level Agreement\\
    QoS & Quality of Service\\
    RL & Reinforcement Learning\\
    SecSLA & Security Service Level Agreement\\
    SLA & Service Level Agreement\\
    SLI & Service Level Indicator\\
    SLM & Service Level Management\\
    SLO & Service Level Objective\\
    WSDL & Web Services Description Language\\
    WSLA & Web Service Level Agreement\\ 
    WSN &  Wireless Sensor Network\\
  \bottomrule
\end{tabular}
\end{table}

\subsection{SLA Definition}
\label{sub:SLAdefinition}

A Service Level Agreement (SLA) is a formal contract between a service provider and a customer that outlines the level of service the customer can expect. 
The European Commission Report on recent European and national projects covering Cloud computing SLAs \cite{kyriazis2013cloud} gives a formal definition of such a recommendation:

\begin{quote}
   {\em A Service Level Agreement (SLA) is a formal, negotiated document that defines (or attempts to define) in quantitative (and perhaps qualitative) terms the service being offered to a Customer. Any metrics included in an SLA should be capable of being measured on a regular basis and the SLA should be recorded by whom.} 
\end{quote}

\noindent
According to the previous definition, the purpose of an SLA is to define the terms, conditions, and expectations for the service being provided, as well as the responsibilities of both parties \cite{wustenhoff2002service,papadopoulos2017slas}.
Since SLAs provide a clear and measurable framework for understanding and managing service delivery, they represent key factors in Service Level Management (SLM), which is the process that focuses on managing and improving the quality of services provided by an IT service provider.

Specifically, in \cite{papadopoulos2017slas} several 
characteristics defining a proper SLA are listed, namely:

\begin{itemize}
    \item {\bf Attainability} is the possibility of meeting the desired level of service;
    \item {\bf Meaningfulness} is a property defining that all SLA parts must be relevant to the agreement;
    \item {\bf Measurability} defines that the level of service provisioning should be measurable in an impartial way;
    \item {\bf Controllability} specifies that the factors impacting the SLA must be under the service provider's control;
    \item {\bf Understandability} means that both parties must understand the concepts and quantities of the SLA;
    \item {\bf Affordability} is a property determining that the SLA should be cost-effective;
    \item {\bf Mutual acceptability} is related to the definition of the SLA that should be the result of the negotiation between parties.
\end{itemize}

As an ``agreement'', an SLA includes a set of different features regarding the provisioning of the service. These refer to the agreed Quality of Service (QoS, hereafter) declined through different terms, the Service Level Objectives (SLOs), the responsibilities and obligations of the parties, as well as the penalties in cases of non-compliance to the agreed terms. Specifically, the following components must be present in an SLA \cite{jin2002analysis,girs2020systematic}:

\begin{itemize}
    \item {\bf Service Name and Description} provide the name and description of each offered service and the main objective of the whole SLA.
    \item {\bf Parties} describe an individual or group of entities involved in the contract and their roles (i.e., service provider and consumer). A party could be a private, commercial, or public entity.
    \item {\bf Validity Period} defining the period covered by the SLA.
    \item {\bf Scope} of the agreement.
    \item {\bf Restrictions}, defining the steps to be taken to provide the service.
    \item {\bf Service Level Indicator} (SLI) is a parameter or a metric associated with a service able to specify a certain quantitative or qualitative level of service. Examples include availability, latency, response time, jitter, scalability, processing capacity, memory, storage, and so on.
    \item {\bf Service Level Objective} (SLO) representing a threshold or a value on an SLI or a metric.
    \item {\bf Penalties} defining what happens in case the service provider is unable to meet the objectives in the SLA. Specifically, the penalty parameter refers to the fee that the service provider must pay to the service consumer if the SLO specified on an associated SLI is not met. 
    \item {\bf Optional services} which are not mandatory for the user.
    \item {\bf Exclusion} parameters are elements specifying what is not covered in the SLA.
    \item {\bf Administration} describing the processes created in the SLA to meet and measure its objectives and defines organizational responsibility for the monitoring of those processes. 
\end{itemize}

Moreover, according to the different perspectives an SLA may have and the services offered, two main SLA modalities exist \cite{uriarte2019defining}. The former type is {\em Service-Based} SLA, which refers to non-negotiable and ready-to-sign SLAs available for all consumers. This is the common type of SLA in Clouds, implying that the service agreements are the same across all customers.
The latter is {\em Customer-Based} SLA, which has negotiable terms and is agreed upon with individuals or groups to adapt the services to their needs. Although more flexible, this modality is significantly more complex and less used.

\subsection{SLA Life cycle}
\label{subsec:lifecycle}

\begin{figure}
    \centering
    \includegraphics[scale=0.6]{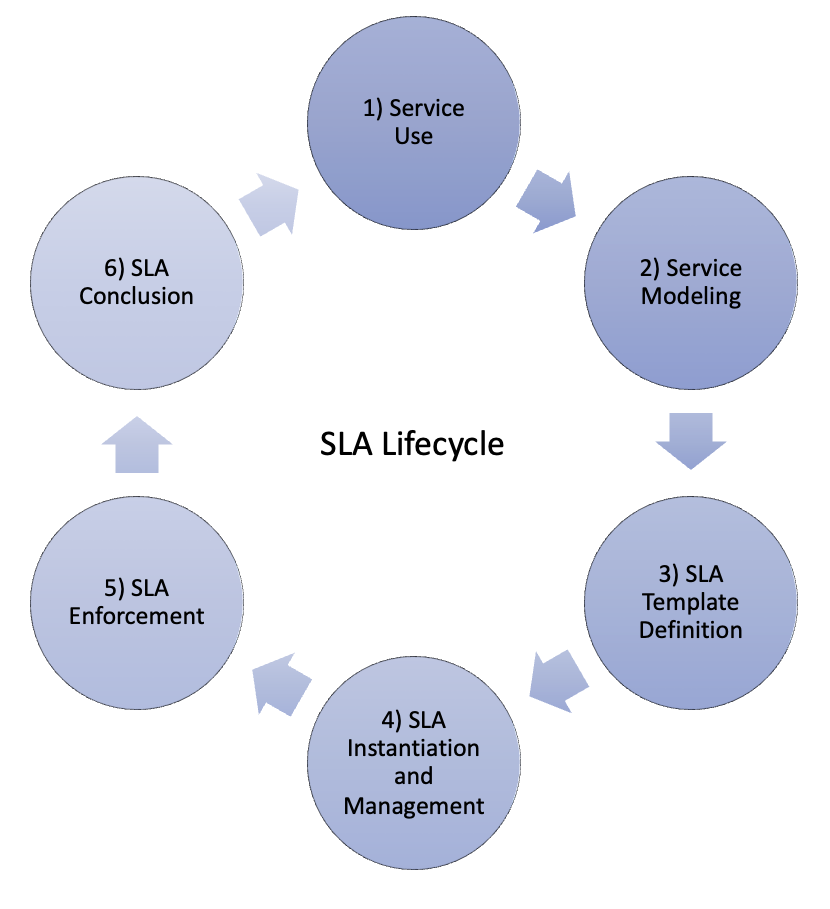}
    \caption{SLA life cycle}
    \label{fig:SLALifecycle}
\end{figure}

The SLA life cycle meta-model was first discussed in the European Commission Report \cite{kyriazis2013cloud}. As shown in Figure \ref{fig:SLALifecycle} it captures the main phases, structures, processes, and entity interactions in the SLA life cycle. 

The six main phases of the SLA life cycle include:

\begin{itemize}
    \item {\bf Service Use}. This phase refers to the SLA definition. Before the services can be provided to the consumer, both the provider and the consumer must agree on the terms of the agreement, such as metrics, level, quality, price, and penalties. 
    \item {\bf Service Modeling}. This phase deals with the modeling of the service, relationships, dependencies within the service components, and information regarding the service provision. The outcome of the process could be an artifact or document (in a standard modeling language, such as XML), which includes all the parameters affecting the service execution, usage, and delivery. 
    \item {\bf SLA Template Definition}. In this phase, SLA templates are created and other related information is captured. 
    \item {\bf SLA Instantiation and Management}. This phase deals with the mechanisms for discovering providers for specific services and dynamic negotiation between participating entities.
    \item {\bf SLA Enforcement}. This phase retains the quality parameters (agreed in signed SLAs). All providers exploit monitoring mechanisms and SLA violation detection mechanisms to trigger corrective actions.
    \item {\bf SLA Conclusion}. This phase handles the termination of the SLA, which can happen for various reasons such as the service delivery has been successfully concluded, the SLA validity period has expired, or an SLA violation has occurred.
\end{itemize}

Starting from the above SLA life cycle and the technical classification used in \cite{mubeen2017management}, we identify a criterium to categorize the analyzed literature and group the different studies in the rest of this investigation. Specifically, we leverage the following classification categories:

\begin{itemize}
    \item {\bf SLA Modeling}. This category comprises frameworks, templates, and modeling languages to model SLA solutions that have been proposed in the recent literature. This category can be mapped to the SLA Modeling and SLA Template Definition stages of the SLA life cycle.
    \item {\bf SLA Negotiation}. This category includes papers focusing on group framework to create non-negotiable, negotiable, and re-negotiable SLAs. This group concerns the SLA Instantiation and Management phases of the SLA life cycle.
    \item {\bf SLA Monitoring and Violation}. Approaches to check and evaluate the expected level of service are grouped in this category. Moreover, papers on the economic penalties derived from possible SLA violations are also analyzed.
    This category can be mapped to the SLA enforcement and SLA Conclusion stages of the SLA life cycle.
\end{itemize}

Table \ref{tab:mappingSLALifecycle} shows the mapping between the categorization strategy used throughout this article and the different stages of the SLA life cycle. Observe that, not all the SLA life cycle categories are used in this paper to categorize the literature research. In particular, the Service Use stage has been mainly discussed in very old research proposals and, hence, we included them in Section \ref{sub:SLAdefinition}, where we defined background concepts related to SLA and its characteristics.

\begin{table}
\centering
  \caption{Mapping between categorization used throughout this paper and the different stages of the SLA life cycle\label{tab:mappingSLALifecycle}}
  \begin{tabular}{ll}
    \toprule
    SLA life cycle Stage & Our Categorization Strategy\\
    \midrule
    1. Service Use & SLA Definition (see Section \ref{sub:SLAdefinition})\\
    2. Service Modeling & SLA Modeling (see Section \ref{sub:SLAmodeling})\\
    3. SLA Template Definition & SLA Modeling (see Section \ref{sub:SLAmodeling})\\
    4. SLA Instantiation and Management & SLA Negotiation (see Section \ref{sub:SLAnegotiation})\\
    5. SLA Enforcement & SLA Monitoring and Violation (see Section \ref{sub:SLAmonitoring})\\
    6. SLA Conclusion & SLA Monitoring and Violation (see Section \ref{sub:SLAmonitoring})\\
  \bottomrule
\end{tabular}
\end{table}

\subsection{SLA Language Specifications}
\label{sub:SLALanguage}

This section discusses the state-of-the-art research on SLA
specification languages.
The procedure through which different SLAs are automatically described, provisioned, and observed is quite recent. Before it, SLA contracts were mostly written
using natural expressions, and the examination of compliance was done manually \cite{keller2003wsla}. The current most prominent industrial approaches for SLA language specification are:
WSLA, WS-Agreement, SLA$*$, CSLA, SLAC, RBSLA, and SLA-IoT \cite{maarouf2015review}.

\begin{figure}
    \centering
    \includegraphics[width=0.6\linewidth]{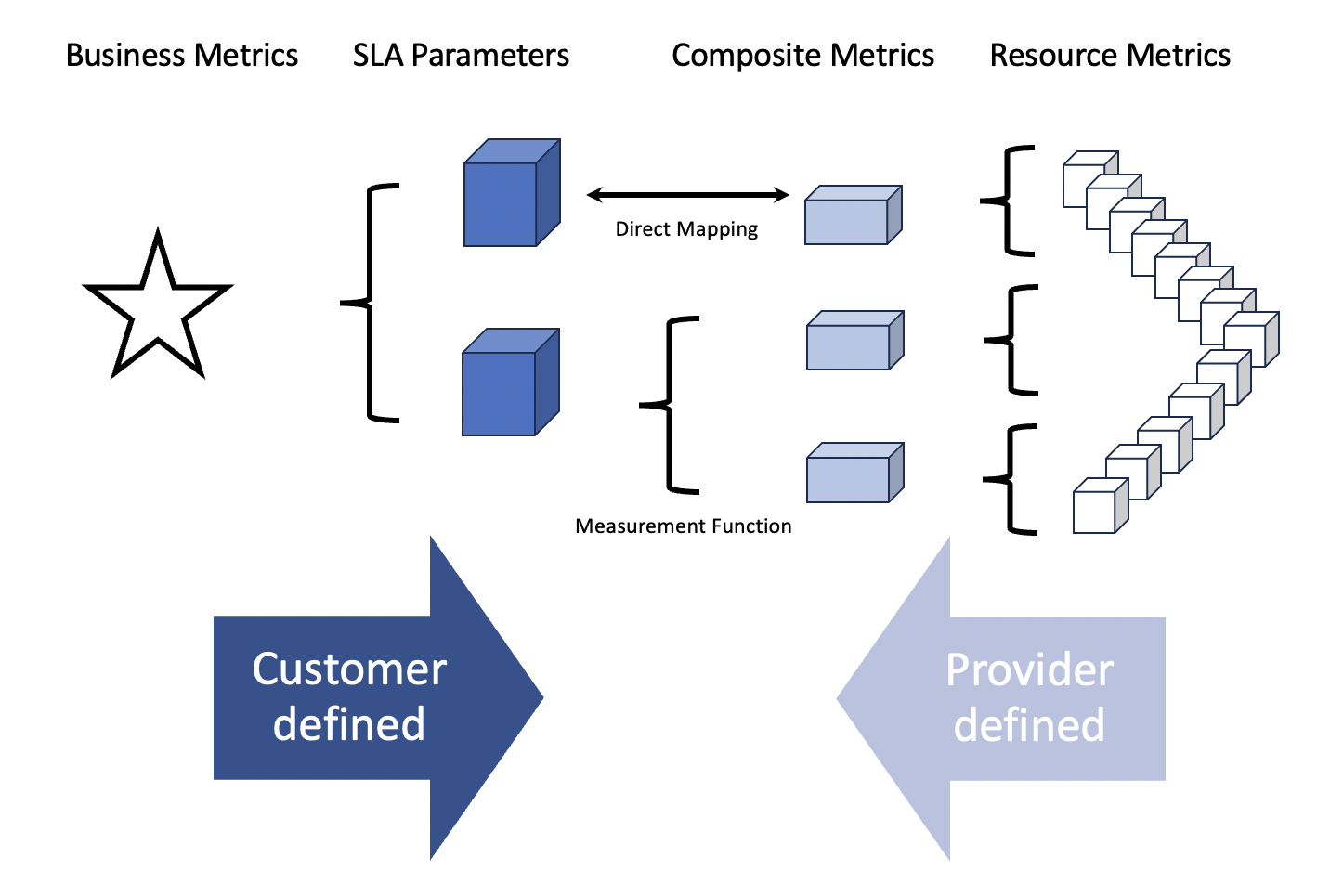}
    \caption{WSLA terminology and main components}
    \label{fig:WSLA}
\end{figure}

{\bf WSLA.} IBM published the Web Service Level Agreement (WSLA),
which provides a specification for the definition and monitoring of SLAs within a Web Service environment \cite{ludwig2003web}. A WSLA provides a runtime architecture and a language for SLA
specification for Web services documents. It describes the assertions of a service provider concerning the service 
including {\em (i)} the agreed parameters (e.g., response time and throughput) derived from business metrics and {\em (ii)} the measures to be taken in case of
failure to meet the service guarantees (for instance, a notification of the service customer). 
As shown in Figure \ref{fig:WSLA}, the service provider's assertions are based on a detailed definition of both
basic and composite service metrics. Underlying, these are derived from low-level metrics related to the resources residing in the service provider tier.

In addition, a WSLA also expresses the actor included in the architecture, namely the monitoring party, third parties that contribute to the measurement of metrics, and the management of deviations of service guarantees. 
The different parties define a proprietary implementation policy for a service that has to be translated into system-level configuration information, thus creating an independent deployment function that interprets the WSLA and takes
appropriate actions for each party.

The WSLA language is based on XML and defined as an XML schema. 
The language is extensible to include: {\em (i)} specific types of operation descriptions (e.g., using WSDL to describe a Web services operation), {\em (ii)} measurement directive types for specific
systems, {\em (iii)} special functions to compose aggregate metrics, and {\em (iv)} predicates to evaluate specific metrics. 

Observe that, a WSLA defines the agreed performance characteristics and how to evaluate and measure them. Instead, a service description, such as the Web Services Description Language (WSDL), describes the service interface relationship between a service and its application. 

{\bf WS-Agreement.} WS-Agreement is another XML-based Web Service SLA specification defined by Open Grid Forum (OGF) \cite{andrieux2007web}. The specification is composed of {\em(i)} two schemas for specifying both the agreement and
the agreement template, and {\em(ii)} a set of port types and
operations for managing the different phases of the agreement life cycle (including creation, expiration, and
monitoring of agreement states).
WS-Agreement contains more information about
the service functional's properties than WSLA and it allows for the extension of new domain-specific elements.

{\bf SLA$*$}. An abstract SLA syntax named SLA$*$ is proposed in \cite{kearney2010sla} to automate the Cloud SLA life cycle. SLA$*$ is
a domain-independent syntax for machine-readable SLAs and SLA templates. This solution promotes the formalization of a SLA in any language and for any service, by eliminating the limitation of XML.
This model was designed as part of the FP7 ICT Integrated Project
SLA@SOI\footnote{\url{https://xlab.si/research/finished-projects/slasoi/}}, and has been used in several industrial
use-cases, such as Enterprise IT, live-media
streaming and health-care provision.

{\bf CSLA.} Cloud Service Level Agreement (CSLA) \cite{kouki2012csla} is an SLA language, based on WSLA and SLA$*$, designed to express both SLA specifications and address SLA violations in the context of Cloud services \cite{kouki2014language}. Besides the standard formal definition of contracts comprising validity,
parties, services definition, and guarantees, CSLA is
enriched with features introducing language support
for Cloud elasticity management (i.e., QoS/functionality degradation and an advanced penalty model).

{\bf SLAC.} Like SLA$*$ and CSLA, Service-Level-Agreement for Clouds (SLAC) \cite{uriarte2014slac}
is a domain-specific language that defines SLAs specifically devised for the Cloud
computing domain. It is inspired by the WS-Agreement and
shares many features with the definition and structure of this
language. The metrics available in the language are pre-defined
in light of the requirements of the Cloud domain. The
set of characteristics of the SLAC language, such as multiparty, group definition, and specification of the involved parties
on each term, enables the definition of SLAs
including a broker (i.e., an agent in charge of SLA negotiation). The semantics of the evaluation of an SLA is formulated as a Constraint Satisfaction Problem (CSP) and is intended to verify the agreement consistency and whether the
characteristics of the service are within the established values.

{\bf RBSLA.} Rule-based Service Level Agreement \cite{paschke2005rbsla} follows a knowledge-based approach and uses RuleML \cite{boley2001design} to specify a SLA. This formal-logical and rule-based approach provides a means
to describe complex contract structures as modular rule sets, which can be collaboratively maintained, interchanged automatically, executed, and enforced by a rule engine.

{\bf SLA-IoT} \cite{alqahtani2016end} is a recent end-to-end SLA specification language designed for the Internet of Things. It presents a new layered syntax for specifying SLA for IoT devices and their connections. 
A contract is composed of the classical elements such as parties, SLOs, workflow activities, services, and
infrastructure resources; however, although this language allows for a more precise definition of the user's requirements, it does not specify the penalties, devices, and user preferences.

\section{Service Level Agreement}
\label{sec:SLA}

This section aims to survey the current solutions for SLA management. We grouped the papers according to our classification criterium (see Section \ref{subsec:lifecycle}), which is focused on the SLA life cycle. In particular, to organize our literature analysis, we considered four main topics: SLA modeling, negotiation, monitoring, and violation phases.

\subsection{SLA Modeling}
\label{sub:SLAmodeling}

This category includes recently published works focusing on the definition of ontologies, templates, and languages to model SLAs. 

In the context
of SLA, ontology refers to a method of providing 
a taxonomy, that is a formal, syntactic, and semantic description model of concepts, properties, and relationships between concepts for SLA content \cite{girs2020systematic}. Several studies have been focused on the definition and application of ontologies for SLA \cite{garcia2017modeling,labidi2018cloud,labidi2017cloud}.
The proposal of \cite{garcia2017modeling} introduces an
extension to the Linked USDL family of ontologies called Linked USDL Agreement, a semantic model to specify, manage, and share SLA descriptions on the Web. It relies on Linked Data principles \cite{bizer2011linked} to share and interlink service agreements over the Web. Moreover, it exploits IDEAS, a software simulation framework-based tool, to support or evaluate the presented methods. This software provides a generic online development environment for domain-specific languages (DSL).
Both the contributions presented in \cite{labidi2017cloud} and \cite{labidi2018cloud} rely on the Ontology Web Language (OWL)\footnote{\url{http://www.w3.org/TR/owl-ref/}} to describe classes, constraints, and properties of an SLA. Their approach provides a system to calculate service price, penalty
amount, and violation number, and to verify the SLA validity by defining its termination terms. The main difference is that the system in \cite{labidi2017cloud} specifically focuses on SLA monitoring to prove possible violations.

Several publications, rather than designing an ontology to define and model SLAs, mainly delve into presenting a new language, a language extension, or a grammar to accomplish SLA modeling \cite{engel2018ysla,alqahtani2018end,alqahtani2019service,noureddine2021ml,scoca2017smart,uriarte2019defining}.
For instance, the authors of \cite{engel2018ysla,scoca2017smart,uriarte2019defining} focus on SLA modeling for Web Services and the Cloud. As a matter of fact, the proposal of \cite{engel2018ysla} introduces the {\em ysla} language and engine. This language, based on YAML, defines novel constructs for configurable and reusable building blocks (namely ``Templates'' and ``Scopes''), allowing the identification of the semantic categorizations of observation data together with metrics
parameterizations. Moreover, a prototypical
implementation, called ysla Engine, is illustrated
in a multi-client, multi-provider environment.
Instead, in \cite{scoca2017smart} Scoca et al., define a formal language for
modeling the interaction of offers and requests. They rely on dSLAC \cite{uriarte2016dynamic}, a language that can be used for specifying
smart contracts in the Cloud domain, and provide evidence that
such contracts have clear advantages. 
Analogously, the work in \cite{uriarte2019defining} introduces a novel SLA definition language, called SLAC, specifically devised for Cloud systems as a formalism to support the whole SLA life cycle. It relies on a mechanism for dealing with the dynamicity of SLA
terms, and to allow the elasticity of Cloud services and guarantee flexibility to the involved parties.
SLAC supports multiple parties and all the roles in the Cloud
service provision, it permits dynamic service level modifications
according to pre-defined conditions and is equipped with a set
of software tools supporting SLA management. A limitation of this approach is that it does not allow for the specification of device and user preferences.

Instead, the authors of \cite{alqahtani2018end,noureddine2021ml} still propose grammars and languages for SLA modeling but they focus on the IoT application scenario.
In particular, Alqahtani et al. \cite{alqahtani2018end} design an SLA grammar that considers workflow activities and the multi-layered nature of IoT
applications. They developed a tool for SLA specification
and composition that can be used as a template to generate SLAs
in a machine-readable format and to provide a fine-grained level of user requirement specifications. Moreover, a rule-based recommendation
system called IoT-CANE is intended to automatically suggest configuration knowledge artifacts to multiple layers, required for users during the IoT resource configuration management processes.
To evaluate user satisfaction with IoT-CANE,
the authors conduct a user study with domain experts.
In the extension presented in \cite{alqahtani2019service} the authors propose an enhanced version of the same grammar that achieves $91.70\%$ of generality and $93.43\%$ of expressiveness and it has been evaluated through a GQM approach.
In the proposal of \cite{noureddine2021ml} ML-SLA-IoT, a framework for IoT SLA specification and monitoring is presented.
The language specification part exploits the functionalities of the Domain-Specific Language (DSL) technique. A DSL is a programming domain-specific language designed to
provide solutions to the problems raised in a particular field. ML-SLA-IoT describes QoS levels, provided
services, and obligations. Moreover, compared to other SLA specification languages, it allows the specification of user preferences, reuse of
microservices, and use of ML-SLA (multi-level metrics and QoS
according to existing constraints and user preferences). 

Differently from the above works, in \cite{kapassa20185g} the authors try to map high-level requirements,
expressed by users in SLAs, to low-level network
parameters included in policies. Their platform, called 5GTANGO, includes an SLA Generator that creates both the SLA templates for the Service/Infrastructure provider and the final SLA itself.

In Table \ref{tab:SLAModeling}, we report a summarized view of the publications related to the modeling phase of the SLA life cycle, identifying the type of modeling (i.e., ontology, language, grammar or template), the method used to validate the approach, and the application domain.

\begin{table}
\centering
  \caption{Publications related to SLA modeling\label{tab:SLAModeling}}
  \begin{tabular}{lllp{2.7cm}l}
    \toprule
    Paper & Year & Modeling Type & Validation & Domain\\
    \midrule
    Garc{\'\i}a, et al. \cite{garcia2017modeling} & 2017 & Ontology & Custom tool & Cloud\\
    Scoca, et al. \cite{scoca2017smart} & 2017 & Language & IDEAS software & Web Service\\
    Labidi et al. \cite{labidi2017cloud} & 2018 & Ontology & Cloud SLA Monitoring Ontology (CSLAMOnto) & Cloud\\
    Kapassa et al. \cite{kapassa20185g} & 2018 & Template & - & 5g Network\\
    Labidi et al. \cite{labidi2018cloud} & 2018 & Ontology & Cloud SLA Analyzing (CSL2A) & Cloud\\
    Engel et al. \cite{engel2018ysla}& 2018 & Language & ysla Engine & Cloud\\
    Alqahtani et al. \cite{alqahtani2018end,alqahtani2019service} & 2018, 2019 & Grammar & Human experts & IoT\\
    Uriarte et al. \cite{uriarte2019defining} & 2019 & Language & Custom tool & Cloud\\
    Noureddine et al. \cite{noureddine2021ml} & 2021 & Language & DSL & IoT\\
  \bottomrule
\end{tabular}
\end{table}

Several approaches do not specifically design a modeling method but propose approaches supporting this phase of the SLA life cycle \cite{ganapathy2022semantically,longo2018comparing}.
In this sense, a further step has been conducted by Ganapathy and Joshi \cite{ganapathy2022semantically}. Here the authors develop a framework to automate the process of extracting knowledge embedded in Cloud SLAs and representing it
in a semantically rich knowledge graph. In this case, the textual SLA in input is processed with the help of Semantic Web technologies and text mining techniques. Moreover, the extracted components are, then, stored through the usage of OWL and RDF graphs. 

Finally, another research still related to the proposals described in this section but not directly focusing on SLA modeling has been carried out in \cite{longo2018comparing}. In particular, SLA elements are conceptualized to support their analysis and comparison through an ontological representation. The proposed ontology focuses on SLA indicators and technical parameters and is built by exploiting the OWL2-RL language.

\subsection{SLA Negotiation}
\label{sub:SLAnegotiation}

In this section, we review research works proposing approaches for SLA negotiation. SLAs are defined before service usage and are based on a negotiation stage between provider and consumer. This phase can happen autonomously provider-side
within any number of constraints, or can be performed interactively, thus supporting more dynamic service environments. In any case, it aims at maximizing revenues while minimizing the service cost to both parties, realigning the delivered service with the current business strategies. 
In most of the existing negotiation protocols, SLA templates or profiles
are offered by the service provider, either represented by humans or by an SLA manager service, to initiate the negotiation. 
Negotiation between customers and providers is usually achieved through a
negotiation protocol. 

Two key approaches can be used during this phase of the SLA life cycle, namely {\em bargaining-based} negotiation (or {\em brokering}) and {\em offer–based} negotiation \cite{hani2015renegotiation}. In the first case, SLA managers, acting as brokers, leverage the service offering or catalog model for customers who choose services based on their preferences. In this type of negotiation, both parties can make offers and counteroffers until an agreement is reached. In the latter case, instead, the SLA manager helps the provider to allocate the resources as needed, providing some offers with various SLO levels to achieve the expected requirements.

A further categorization for this phase is 
based on the number of parties involved. Negotiation can be classified as {\em bilateral} (or one-to-one) and {\em multilateral} (one-to-many). Bilateral negotiation is performed directly and it is one of the most widely used negotiation styles.
Multilateral negotiation, instead, is used in Cloud environments, where providers compete with one another to attract customers. Usually, it takes place when a customer sends service requirements to a negotiation broker to find the most suitable service from providers at a low cost.

A subsequent process, especially useful in Cloud environments and known as renegotiation, is also available to alter the terms of an existing agreement. In the SLA life cycle, it is placed after the SLA Monitoring stage and allows customers and providers to initiate changes in an established agreement before service/resource reconfiguration.

Renegotiation can be proactive or reactive.
In the first case renegotiation can happen in two ways, namely:

\begin{itemize}
    \item the customer begins the renegotiation to support the changing requirements before service termination;
    \item the provider initiates the renegotiation to prevent any violation of the agreed-on service level due to some unexpected environmental change.
\end{itemize}

In the reactive case, instead, when any SLA violation is detected, renegotiation takes place instead of the termination of the service.

In the following, we review several contributions related to SLA negotiation and renegotiation phases. 

The papers described in \cite{li2019model,haddar2020service,scoca2017smart,kumar2023model,rajavel2021cloud,kumar2023user} propose negotiation mechanisms that use distributed service brokers to dynamically negotiate with service providers on behalf of service consumers.
The model described in \cite{li2019model}, called IoT-Negotiate, relies on a hierarchical topology to address the communication challenges in an IoT environment, performs location-based data distribution and replication to enable efficient message forwarding, and conducts distributed SLA negotiation with candidate service providers.
The proposal of \cite{haddar2020service}, instead, relies on a Multi-Criteria Decision Making (MCDM) method to maximize a utility function so that the customer can choose services with the required QoS performances. Also, a negotiation model for the SLA and a context-based SLA contract ontology in IP Multimedia Subsystem (IMS) is designed.
The work in \cite{scoca2017smart} proposes a methodology for SLA autonomous negotiation. The presented framework checks the consistency of the specification, analyzes the
compatibility between offers and requests and finds the best
possible agreement (if any) by leveraging utility functions and
the level of flexibility offered by providers and consumers.

Similarly, more recent works  \cite{kumar2023model,ibrahim2023design,kumar2023user,shojaiemehr2019three,rajavel2021cloud,li2018agent} illustrate negotiation frameworks in Cloud relying on intelligent agents as third parties. In particular, \cite{kumar2023model} describes an Intelligent Automated Negotiation Agent (IANA) designed to meet the dynamic needs of compound services in a Cloud environment to negotiate SLA agreements. 
In \cite{rajavel2021cloud}, an automated dynamic SLA negotiation framework, called ADSLANF, is proposed. It uses a dynamic SLA concept to negotiate service terms and conditions through negotiation mechanisms based on game theory and a third-party broker agent.
An Intelligent Recommender and Negotiation
Agent Model (IRNAM) is illustrated in \cite{kumar2023user}. Firstly, it aims to recommend and select matching Cloud server providers and, then, it finalizes
the SLA to achieve consensus based on the satisfaction level
for all the parties. Because the recommendation phase deals with the choice of a particular provider, the negotiation style is bilateral.

The authors of \cite{di2017supporting} adopt a different perspective allowing users in a multi-cloud environment to express their requirements
in Cloud plan selection. This facility, which is a particular kind of service negotiation, is also known as ``services selection'' and it is carried out using SLA. After identifying user requirements, they illustrate possible approaches to rank Cloud
plans based on users' preferences according to loud exposed SLA.
This approach can be used in
brokerage services, to evaluate offers by different providers,
as well as in single-provider scenarios, to evaluate the different plans (possibly resulting from customization) offered
by a provider. Also, the approaches presented in \cite{taha2017sla,di2019fuzzy} select the
optimal combination of services, from multiple Cloud providers, that best satisfy the customer requirements, relying on exposed SLA. In particular, the approach in \cite{di2019fuzzy} leverages a fuzzy-based brokering service for Cloud plan selection that allows users to specify their requirements using natural language expressions. Fuzzy logic and fuzzy inference systems are adopted
to assess the compliance of Cloud services quantitatively expressed through SLA and hence help users in the Cloud service selection process.

Several contributions rely on {\em offer–based} negotiation \cite{uriarte2019defining,shojaiemehr2019three,li2018agent,li2021trust}.
The authors of \cite{uriarte2019defining} introduce SLAC, an SLA definition language for Cloud environments. SLAC supports the SLA life cycle and the definition of templates for
the negotiation phase to discover the compatibility of offers and requests. In this paper, the authors do not deep dive into negotiation aspects, but they consider that
all involved parties have a Negotiator component
that proposes SLAs and evaluates requests for both negotiation, renegotiation, or modifications in the SLA.
An Automated Negotiation System (ANS) for SLA negotiation of composite Cloud services is proposed in \cite{shojaiemehr2019three}. It deals with the design and development of a novel negotiation strategy that considers three factors for generating proposals, namely: time, negotiators' preferences, and a negotiation method called opponent's behavior.
This method attempts to generate a proposal
that provides higher utility for the customers relying on their proposals.
The paper \cite{li2018agent} takes a different direction by designing an Agent-based Fuzzy Constraint-directed Negotiation (AFCN) model for SLA negotiation that supports an iterative many-to-many infrastructure that does not require a broker to coordinate the negotiation process. To consider the behavior of different agents, this model can also adopt different negotiation strategies, such as competitive, win-win, and collaborative strategies in different Cloud computing environments. 
Similarly, in \cite{li2021trust,li2019tslam}, the authors address the problem of SLA negotiation candidates
selection in IoT through a trust model designed to help Cloud entities make service decisions. In particular, in \cite{li2021trust}, they design this trust model to identify trusted service providers before attempting to negotiate an SLA. The solution leverages both Rough Set theory to predict the negotiation success rate, and Bayesian inference to deduce the possibility of SLA violations according to the monitored data. By contrast, the work presented in \cite{li2019tslam} equips brokers with a learning module enabling them to capture implicit service demands and find user preferences. 

Differently from the previous proposals, the authors of \cite{franceschetti2019checking}
focus on a particular problem related to SLA negotiation. Specifically, they design an approach to allow service providers to check, at design time, whether a composed Web service with temporal parameters will always satisfy the temporal
constraints specified in an SLA.

The SLA renegotiation phase is considered in \cite{paputungan2018real,labidi2020cloud}.
In particular, the paper presented in \cite{paputungan2018real} introduces
an extension of the SLA management life cycle for enabling SLA renegotiation during service delivery. In particular, a real-time and proactive SLA renegotiation model for
dynamic Cloud-based environments is proposed. To achieve real-time decisions, a multi-offer generation approach is
used and a mechanism to detect and predict service violations
is used to ensure proactive renegotiation. 
The strategy proposed in \cite{labidi2020cloud}, instead, models both negotiation and renegotiation phases.
The style of negotiation considered is multilateral, indeed, in this approach there are several participants, both vendors offering their products and buyers submitting bids. Moreover, it relies on an ontological representation to semantically represent the client's requirements and the provider's offers.
The renegotiation stage is automatically triggered during runtime once an unexpected variation in the Cloud service context is detected.

In Table \ref{tab:SLANegotiation}, we report a summarized view of the publications related to the negotiation phase of the SLA life cycle, identifying the type of service (i.e., brokering or offering), the style of negotiation related to the number of parties involved (i.e., bilateral or multilateral), the type of renegotiation if considered in the paper (i.e., proactive or reactive), and 
the application domain.

\begin{table}
\centering
  \caption{Publications related to SLA negotiation\label{tab:SLANegotiation}}
  \begin{tabular}{p{3.4cm}lp{1.5cm}p{2.2cm}p{2.2cm}l}
    \toprule
    Paper & Year & Service Type & Negotiation style &  Renegotiation type & Domain\\
    \midrule
    Scoca et al. \cite{scoca2017smart} & 2017 & Brokering & Multilateral & - & Cloud\\
    Di Vimercati et al. \cite{di2017supporting}& 2017 & Brokering & Multilateral  & - & Cloud\\
    Paputungan et al. \cite{paputungan2018real} & 2018 & - & - & Proactive & Cloud\\
    Li et al. \cite{li2018agent}& 2018 & Offering & Multilateral & - & Cloud\\
    Li et al. \cite{li2019model} & 2019 & Brokering & Multilateral & - & IoT\\
    Uriarte et al. \cite{uriarte2019defining}& 2019 & Offering & Bilateral  & Proactive & Cloud\\
    Li et al. \cite{li2019tslam} & 2019 & Brokering & Bilateral & - & Cloud\\
    Di Vimercati et al. \cite{di2019fuzzy}, Shojaiemehr et al. \cite{shojaiemehr2019three} & 2019 & Brokering & Multilateral & - & Cloud\\
    Labidi et al. \cite{labidi2020cloud}  & 2020 & Offering & Multilateral & Reactive & Cloud\\
    Haddar et al. \cite{haddar2020service} & 2020 & Brokering& Multilateral & - & IMS\\
    Li et al. \cite{li2021trust}& 2021 & Offering & Bilateral  & - & IoT\\
    Rajavel et al. \cite{rajavel2021cloud}& 2021 & Brokering & Multilateral & - & Cloud\\
    Kumar et al. \cite{kumar2023model}, Ibrahim et al. \cite{ibrahim2023design} & 2023& Brokering & Multilateral  & - & Cloud\\
    Kumar et al. \cite{kumar2023user}& 2023& Brokering & Bilateral & - & Cloud\\
  \bottomrule
\end{tabular}
\end{table}

An alternative approach to renegotiation is presented in \cite{grubitzsch2017ml} and it is called Multi-Level SLA (ML-SLA). ML-SLA supports the dynamic change of service
levels with price adjustments to overcome the limited flexibility issues coming from the traditional static SLA approaches. Compared to traditional SLA
approaches with renegotiation, in the Multi-Level SLA approach services do not have
to be terminated or interrupted and level switches require less time and effort.

\subsection{SLA Monitoring and Violation}
\label{sub:SLAmonitoring}

The monitoring process in the SLA life cycle aims to capture the performance of physical and virtual servers continuously, the network, the shared resources, and the applications running on top of them \cite{hani2015renegotiation}.

Monitoring can be used to detect whether an SLA has been violated.
Predicting possible SLA violations can be useful in proactively maintaining the SLAs. By monitoring resource utilization, the service provider can study past disappointments and avoid them in the first occurrence.

When the service lifetime is expiring or when any unacceptable violation has occurred, the contract of service termination would be initiated by either the service provider or the customer. In the latter
case, the service provider incurs a penalty. 
Specifically, SLA violations can be due to one of the following reasons\cite{rana2008monitoring}:

\begin{itemize}
    \item Defective performance, i.e., the monitored parameters have lower levels.
    \item Late performance, the service is provided at the appropriate level but with unjustified delays.
    \item The service is not provided at all.
\end{itemize}
Any type of breakdown has consequences regarding customer satisfaction and may result in a violation of an SLA \cite{alodib2016qos}.

In the following, we describe recent approaches defining solutions for SLA monitoring, violation, or prediction of violation.

In particular, the proposals of \cite{prasad2020preserving,badshah2023service,prasad2020monitoring} focus on SLA monitoring. The authors of \cite{prasad2020preserving} describe a threshold method for preserving SLA parameters for Trusted IaaS Cloud. They identify particular SLA metrics to monitor to assess the status of the service provisioning. If they recognize that the value is less than a threshold value, then precautionary actions are taken to avoid breaching the SLA.
In \cite{badshah2023service}, the authors propose SLA-Monitoring as a Service (SLA-MaaS), a framework for monitoring providers' services by adopting third-party monitoring services with SLA and penalties management. The violation part is 
handled via a three-layer penalties approach, i.e., if a violation occurs, the system starts from a warning and applies lower penalties, instead of directly terminating the SLA. 
In \cite{prasad2020monitoring}, the authors describe a mechanism for SLA monitoring
using Reinforcement Learning (RL) and a Long short-term memory (LSTM) network for the prediction of Cloud resources.

The authors of \cite{singh2023autonomous,mahan2021novel,karamanlioglu2018ontology} propose approaches dealing with SLA violations.
In particular, Singh and Goraya \cite{singh2023autonomous}
propose an automatic multiagent framework that ensures the minimization of the SLA
violation rate in workload execution. A negotiation agent selects the best service that can execute the current workload
without SLA violation and with minimum consumption of
energy. A monitoring agent has to
continuously monitor for SLA violations, whereas another agent keeps track of workload executions
to provide better forecasts of future executions. 
Similarly to this approach, \cite{mahan2021novel} presents a method that identifies the physical hosts' workloads before the overflow energy consumption in a Cloud environment, while also reducing SLA violation. This approach predicts the load of physical hosts through both an energy-conscious model and a granular neural network. 
In \cite{karamanlioglu2018ontology}, a generic SLA ontology is created to develop an expert system called SLAVIDES using it. This framework aims to detect SLA violations, check constraints, and make inferences. Since the ontology is designed as generic, it is intended to be used in many different knowledge domains.

Differently, the proposal of Uriarte et al. \cite{uriarte2019defining} defines SLAC, as an SLA definition language for Clouds to support the whole SLA life cycle, including the monitoring phase. The penalty and billing enforcement module is invoked only after termination, hence, the violation phase is not included in the described module. To perform monitoring, SLAC uses a component called
SLA Inspector parses the data received from the Monitor module and generates a set of constraints. The satisfiability of these constraints is, then, checked against the SLA constraints using the Z3 solver \cite{de2008z3}, a Satisfiability Modulo Theories (SMT) solver from Microsoft Research.

Other more recent studies \cite{swain2023key,subeh2021learning,zeng2020detection} apply Machine Learning (ML) models to a large IT service dataset to predict whether an incident involves a violation of SLA conditions. In particular, \cite{swain2023key} makes a comparison of ML models and applies them to a large IT services incident dataset. The authors found that logistic regression and neural network models have the best performance in terms of misclassification rates and average squared error. This investigation can be useful for proactive incident management.
The work in \cite{zeng2020detection} explores four diverse ML-based
predictors (i.e., logistics regression, artificial neural network,
random forest, and extreme gradient boosting) to detect and predict SLA violations.
Instead, \cite{subeh2021learning} exploits an ML technique based on the Learning Classifier System (LCS) to detect violations.

A consistent group of proposals adopts Blockchain technology for its potential to support monitoring or violation detection approaches without the use of trusted third parties, guaranteeing the integrity of the client's logs\cite{battula2022blockchain,neeraj2023service,noureddine2021ml,alzubaidi2020blockchain,uriarte2021distributed,ranchal2020slam,pandey2021sla,abhishek2021sla,de2020service}.
In particular, the authors of \cite{neeraj2023service} leverage a public Blockchain and a log-based algorithm to detect SLA violations in the Cloud. SLA parameters are monitored by the service provider using the agreed SLA template.
The proposals of \cite{noureddine2021ml,de2020service,wonjiga2019blockchain}
relies on Blockchain and smart contracts to monitor the SLA terms without the intervention of a third party. 
The authors of \cite{wonjiga2019blockchain} propose to verify the SLA using an integrity-checking method based on a distributed ledger. 
The framework called ML-SLA-IoT \cite{noureddine2021ml} 
automatically generates smart contracts from the SLOs, these
smart contracts are responsible for tracking the Service Level Objective parameters, detecting violations, and notifying the service
provider. 
The proposal of \cite{alzubaidi2020blockchain} consists of a decentralized approach for enforcing the consequences of SLA violations. It relies on smart contracts to address incidents and automate decision-making on the compliance level of providers.
The authors of \cite{uriarte2021distributed} include in their framework a two-level Blockchain architecture. At the first level, the
smart SLA is transformed into a smart contract, based on SLAC\cite{uriarte2019defining}. At the second level, a permissioned Blockchain is in charge of generating objective measurements for the smart SLA/contract assessment. 
Instead, the work of \cite{battula2022blockchain}
focuses on fog computing, while the proposal described in \cite{ranchal2020slam} designs SLAM, a framework for continuous SLA monitoring in a multi-cloud ecosystem. Similar to the above papers, it is based on Blockchain and leverages
smart contracts to detect SLA violations, but it determines the violations' root causes through a hierarchical system structure.
Analogously, the proposal of \cite{abhishek2021sla} relies on a Blockchain network for SLA violation detection. Moreover, it provides an interface for both the Cloud service provider and clients to access and store the violation logs in an immutable system.
The framework illustrated in \cite{pandey2021sla} is composed of an SLA monitoring module developed using the auto-scaling feature of OpenStack Hadoop Service, a violation and logging module implemented through Blockchain. Moreover, the framework includes an algorithm to
compensate the users based on the number of violations and
the type of subscription. 
The proposal of \cite{zhou2018trustworthy} relies on several witnesses to perform SLA monitoring and detect violations. These actors must behave honestly to gain
the maximum profit for themselves, which can be proved by a game
theory strategy. Compensation in case of violation is automatically transferred to the customer leveraging a Blockchain smart contract.

In Table \ref{tab:SLAMonitoring}, we report a summarized view of the publications related to both the monitoring and violation phases, because, as seen before, these two stages of the SLA life cycle are closely linked together. This table shows if the cited paper presents an approach for monitoring, violation, or violation prediction, the technique it applies, and the application domain.

\begin{table}
\centering
  \caption{Publications related to SLA Monitoring and Violation\label{tab:SLAMonitoring}}
  \begin{tabular}{p{2.8cm}llp{1.8cm}p{1.5cm}p{1.5cm}}
    \toprule
    Paper & Year & Monitoring & Violation & Prediction & Domain\\
    \midrule
    Karamanlioglu and Alpaslan\cite{karamanlioglu2018ontology} & 2018 & \cmark(expert system) &- & - & -\\
    Zhou et al. \cite{zhou2018trustworthy}& 2018  & \cmark(witnesses) & \cmark(witnesses)  & - &Cloud\\
    Uriarte et al. \cite{uriarte2019defining} & 2019  & \cmark (Z3 solver) & - & - &Cloud\\
    Wonjiga et al. \cite{wonjiga2019blockchain} & 2019  & \cmark (Blockchain) & - & - &Cloud\\
    Prasad et al. \cite{prasad2020preserving} & 2020 & \cmark (threshold) & - & - & IaaS Cloud\\ 
    Prasad et al. \cite{prasad2020monitoring} & 2020 & \cmark (RL) & - & \cmark (LSTM) & IoT-based Cloud\\
    Alzubaidi et al. \cite{alzubaidi2020blockchain}& 2020 & - & \cmark(Blockchain) & - & IoT\\
    Ranchal and Choudhury\cite{ranchal2020slam} & 2020 & - & \cmark(Blockchain) & - & Multi-Cloud\\
    De Brito et al. \cite{de2020service} & 2020 &  \cmark(Blockchain) & - & - & Cloud\\
    Subeh and Al-Ajeli \cite{subeh2021learning} & 2021 &  - & - & \cmark (LCS) & Business processes\\
    Noureddine et al. \cite{noureddine2021ml}, Uriarte et al. \cite{uriarte2021distributed}& 2021 & \cmark (Blockchain) & - & - & IoT\\
    Pandey et al. \cite{pandey2021sla}& 2021 & \cmark(log analysis) & \cmark (Blockchain)  & - & Cloud\\
    Abhishek et al. \cite{abhishek2021sla} & 2021 & - & \cmark (Blockchain)  & - & Multi-Cloud\\
    Zeng et al. \cite{zeng2020detection}& 2021 & - & \cmark (ML) & \cmark (ML) & Web services\\
    Battula et al. \cite{battula2022blockchain} & 2022& - & \cmark (Blockchain) & - & Fog Computing\\
    Badshah et al. \cite{badshah2023service} & 2023 & \cmark (SLA-MaaS) & \cmark (three-layer penalties) & - & Cloud\\
    Neeraj et al. \cite{neeraj2023service} & 2023 & \cmark (log analysis) & \cmark (Blockchain) & - & Cloud\\
    Singh and Goraya \cite{singh2023autonomous}& 2023 & \cmark (agent) & \cmark (agent) & \cmark (agent) & Cloud\\
    Swain and Garza \cite{swain2023key}& 2023 & - & - & \cmark (ML) & IT services\\
  \bottomrule
\end{tabular}
\end{table}

Finally, we consider the compensation process strictly related to the SLA violation phase.
Indeed, in the last phase of the SLA management life cycle, in case any violation is encountered during the SLA monitoring phase, the compensation process occurs and penalties are enforced. In this direction, the paper presented in \cite{scheid2019enabling} provides a framework for the translation of QoS-related SLAs in Blockchain-based smart contracts as done in \cite{uriarte2021distributed,uriarte2019defining}. Moreover, it automates SLA compensation and service fee payments relying on Blockchain.

\section{Security and Privacy SLA}
\label{sec:secSLA}

In the current Cloud and IoT scenarios, security assurance and
transparency continues to be a significant issue given {\em(i)} the growing number of both devices and Cloud servers offering diverse services, {\em(ii)} new and heterogeneous architectures, and {\em (iii)} the vulnerability to information disclosure and service unavailability of such environments.

Regular SLAs usually concern quantitative and measurable indicators of performance-related Service Level Objectives (SLOs), mainly concerning the availability of service, the response time, and the Quality of Service (QoS). Hence they did not provide coverage of security metrics by definition.

Although security has been considered for decades as a possible attribute in SLAs \cite{henning1999security}, the problem of establishing measurable and shared metrics in this context makes it difficult for researchers and practitioners to design solutions to include security guarantees in their SLAs, explicitly. Hence, finding approaches for managing security in SLA for all the phases of its life cycle (including negotiation, enforcement, and monitoring phases) is still an open challenge and work needs to be done in this direction \cite{casola2015adoption}.

Only recently, Security SLAs (SecSLAs) have been introduced to pave the way for the inclusion of security aspects in service provisioning.
Like the traditional SLA, the life cycle
of SecSLAs comprises several stages each defining a
specific role held by either the customer or the service provider:
\begin{itemize}
    \item {\bf Definition}. This phase includes the specification of the security parameters and metrics possessed by a SecSLA.
    \item {\bf Negotiation}. In this phase, the security requirements are planned by the different actors of the systems. If present, a Cloud broker can initiate its evaluation process. 
    \item {\bf Deployment}. In this phase, the needed security services are deployed through the implementation of security mechanisms.
    \item {\bf Monitoring and Reporting}. After its execution, the SecSLA is continuously monitored. Moreover, during this last phase, multiple actions may take place, namely, {\em (i)} the reporting of security and performance levels, {\em (ii)} the prediction of contract violations, {\em (iii)} the management of corrective actions, and {\em (iv)} the implementation of incident response and remediation plans.
\end{itemize}

To enforce and monitor security in Cloud and IoT environments, a report by the European Union Agency for Network and Information Security (ENISA) recommended the adoption of Security Service Level Agreements (Security SLA). 
According to this report, a Security SLA (SecSLA) is defined as a contract between service providers and service customers stating the level of granted security between the parties\cite{enisa2009enisa,casola2016automatically}. Also, several security metrics have been proposed in the context of European projects
such as A4Cloud\footnote{\url{https://a4cloud.eu/}}, SPECS \cite{casola2013specs}, and MUSA\footnote{\url{https://musa-h2020.eu/}} recently investigating the adoption of
SLAs and Security SLAs in the Cloud by proposing models and
frameworks for SLA definition and life cycle management.
The SPECS project \cite{casola2013specs} is an example of such an effort. Its main purpose is to design
a framework that offers Security-as-a-Service in a Cloud-based environment specifying the security parameters directly in the SLA, thus providing a way to manage its life cycle.
A Security Metric Catalogue including $35$ metrics is integrated into the machine-readable format proposed by the SPECS project for Security SLAs \cite{casola2018security}. These metrics are compliant with the directions from ISO about the structure and format of metrics.

According to Chan et al., \cite{chan2004role}, there are several security properties or dimensions (i.e. availability, data confidentiality, data integrity, access control,
authentication, non-repudiation, communication security, and privacy) from which specific security metrics can be computed and that can be potentially used as Security SLA attributes \cite{nugraha2017understanding}. In particular:

\begin{itemize}
    \item Availability is a property including response and resolution times and is thought to guarantee the lack of denial of service (DOS attack) to the network and the services. The SLA parameter used to measure this metric is the percentage of downtime due to security incidents.
    \item Data confidentiality is intended to preserve sensitive data from unauthorized access or disclosure. 
    \item Data integrity ensures the correctness of data against unauthorized modification or tampering attacks. The percentage of such attacks can provide a metric to measure this property.
    \item Access control guarantees that only authorized personnel or devices are allowed to access network elements, services, and applications.
    \item Authentication is a property applied to identify the communication entities against spoofing.
    \item Non-repudiation property involves the ability to give proof-of-origin to data or the cause of an event or an action. A possible metric for this property can be the percentage of the use of digital signatures.
    \item Communication security ensures that information flows only between authorized endpoints and can be measured through the percentage of hijacked time sessions.
    \item Privacy deals with the protection of information and identity of the involved actors.
\end{itemize}

In the last decade, a further specification of SecSLA has been defined in the context of the Cloud. It is known as the Privacy Level Agreement (PLA, hereafter) and it has been first defined by by the Privacy Level Agreement Working Group of the Cloud Security Alliance (CSA) \cite{CSA2013}. PLAs are similar to SecSLAs, but they deal only with concepts related to data protection practices \cite{d2015towards,diamantopoulou2017privacy}. 

In the following, we start by describing the current research proposing approaches related to SecSLAs \cite{halabi2018broker,teshome2018verification,casola2020novel,rios2022security,nugraha2017towards}.
In \cite{halabi2018broker}, the authors propose a broker-based framework for SecSLA management in the Cloud. They design service selection based on security satisfaction and, to minimize the number of security breaches and incidents, they introduce a SecSLA monitoring model,
a violation prediction, and a remediation process. Moreover, in this work, a set of measurable security metrics is developed.
Another proposal \cite{teshome2018verification} suggests including the security monitoring terms in IaaS Cloud SLAs. Moreover, it implements the monitoring phase of the SLA life cycle by evaluating production traffic in the case of anomaly-based Network IDSs (NIDSs). A continuously observed KPI represented as an SLO, is used in an SLA to verify the SLA itself and describe the performance of an NIDS.
The authors of \cite{casola2020novel} propose a Security-by-Design methodology for the development of secure Cloud applications. This framework relies on SecSLA to specify the application security
capabilities and quantifying the provided level of security.
The adopted Security SLA model has been introduced in the SPECS project \cite{casola2013specs}. It is based on the WS-Agreement's XML schema standard, which
has been extended to include Cloud provider-specific information
and security-related guarantees. 
It aims at implementing a data model to collect security-related information (i.e., assets, threats,
vulnerabilities) and a graph-based model to represent distributed
Cloud applications. Thanks to these, the proposed automated methodology has been developed to support a tool
for {\em (i)} security requirement identification performed by
means of a risk analysis process, {\em (ii)} components (COTS) security
assessments and {\em (iii)} Cloud application security assessment.
The work of \cite{rios2022security} presents a solution for security
SLA creation for health service in multi-Cloud-based IoT including the specification of both security and privacy levels. It presents a methodology to quantitatively calculate security and privacy SLAs of IoT applications on top of standard controls. This is obtained by mapping the application components to the security control
implementation and, then, to the security metrics.
Nugraha and Martin \cite{nugraha2017towards} define Trustworthy
Service Level Agreements (TSLA) as a mechanism for incorporating
privacy parameters (especially confidentiality) into SLAs to preserve the confidentiality of government data against unauthorized access or disclosure. In the paper, they describe five discrete levels of security precautions applied to the formulation of security-related SLA.

A group of works specifically focuses on Privacy SLA (PLA, hereafter)\cite{diamantopoulou2017privacy,zhou2017privacy}, which are specifications of SecSLA related only to privacy metrics.
The study of \cite{diamantopoulou2017privacy} proposes the exploitation of PLAs in the
context of Public Administration's (PA's) Information Systems with several objectives: {\em (i)} define citizens' privacy needs, {\em (ii)} provide feedback on data sharing, and {\em (iii)} enable PA departments to analyze privacy threats and vulnerabilities and compliance with laws and regulations. Instead,  \cite{zhou2017privacy} presents a privacy-based SLA violation detection model
for Cloud computing based on Markov decision process theory. This model can recognize
and handle Cloud providers' actions based on users' requirements. Additionally, the model could evaluate the credibility of Cloud providers, and user privacy violations.

In Table \ref{tab:SecSLA} we report a summarized view of the publications related to the Security and Privacy Service Level Agreement. In particular, this table illustrates if a cited paper is based on a Security (SecSLA) or Privacy Level Agreement (PLA), which phase of the security life cycle is considered, and the paper's application domain.

\begin{table}
\centering
  \caption{Publications related to Security and Privacy SLA \label{tab:SecSLA}}
  \begin{tabular}{p{2.2cm}lllp{3.5cm}p{3cm}}
    \toprule
    Paper & Year & SecSLA & PLA & Life cycle phase & Domain\\
    \midrule
    Nugraha and Martin \cite{nugraha2017towards} & 2017 & \cmark & \cmark & Definition & Government Cloud\\
    Diamantopoulou \cite{diamantopoulou2017privacy} & 2017 & - & \cmark & Definition, Monitoring and Reporting & PA\\
    Zhou \cite{zhou2017privacy} & 2017 & - & \cmark & Monitoring and Reporting & Cloud\\
    Halabi and Bellaiche \cite{halabi2018broker} & 2018 & \cmark & - & Negotiation, Monitoring and Reporting & Cloud\\
    Teshome et al. \cite{teshome2018verification} & 2018 & \cmark & - & Monitoring and Reporting& IaaS Cloud\\
    Casola et al. \cite{casola2020novel} & 2020 & \cmark & - & Definition, Negotiation, Monitoring and Reporting& Cloud\\
    Rios et al. \cite{rios2022security} & 2022 & \cmark & \cmark & Definition, Monitoring and Reporting& Multi-Cloud-based IoT\\
  \bottomrule
\end{tabular}
\end{table}

\section{Open Issues and Future Directions}
\label{sec:future}

After performing our investigation, we have identified various challenges and open issues that can be explored to drive future research in the area. They mainly deal with SLA management, specifically regarding
automation, standardization, integration, security and privacy, scalability, dynamic
changes, and legal regulations \cite{noureddine2021ml}.

\begin{itemize}
\item {\bf Automation of SLA life cycle}. If carried out manually, the steps of the SLA management life cycle, comprising SLA creation, negotiation, monitoring, and violation, may result in
expenses and errors. Guaranteeing that automated systems can interpret and respond appropriately to complex service metrics and contractual obligations is an ongoing challenge. Indeed, researchers and practitioners are endeavoring to develop automated software tools for the management of SLA documents \cite{garcia2017modeling}.

\item {\bf Complexity and standardization}. Existing SLA languages focus on finding common formats or schemas, but often SLAs are technical documents related to terminology and concepts understandable by a specific class of specialists. Therefore, the complexity of services and the lack of standardized SLA frameworks across industries can contrast effective management. This trend is still open and the aim is to build solutions so that services can be described using mutually understandable terms and concepts \cite{garcia2017modeling}. 

\item {\bf SLA integration compatibility}. Especially for multi-cloud or IoT applications, where applications are distributed in diverse infrastructures and deployed in smart devices, SLAs are challenging to agree on due to the different standards adopted by providers, which are often proprietary and unable to interoperate. Ensuring consistency and compatibility across these platforms poses a significant challenge \cite{odun2019cloud,noureddine2021ml}.
Furthermore, for the IoT paradigm, issues of interdependence between devices may exist as the service provided by one of them could depend on the service provided by the others \cite{papadopoulos2017slas,alqahtani2018end}.

\item {\bf Security and Privacy}. Incorporating security measures into SLAs and handling privacy concerns related to sensitive data pose ongoing challenges, yet too little research has been conducted about this topic (see Section \ref{sec:conclusion} for numeric about this concept). This results in a lack of trust among parties about the security capabilities offered by providers. Measuring security capabilities is difficult and, consequently, assessing whether a service provides the same level of security precaution for all customers could be not doable\cite{nugraha2017understanding}. Moreover, even if guaranteeing compliance with evolving data protection regulations is crucial, existing certification schemes are still at an early stage for service provisioning context \cite{nugraha2017towards}. Therefore, research for security metrics development for
SLAs in the current Cloud and IoT environments have proven to be an important topic to be investigated and the establishment of universally accepted key performance indicators (KPIs) in this context should be addressed in the future.

\item {\bf Dynamic and real-time SLA management}. Adapting SLAs to the dynamic nature of current Cloud and IoT paradigms, thus guaranteeing real-time monitoring and adjustments, is still a goal. Systems need to respond promptly to modifications of service conditions and new dynamic negotiation algorithms should be designed \cite{scheid2019enabling}. This is a challenge, especially for Industry 5.0 scenarios, where SLA can be used to trace scalability issues \cite{maddikunta2022industry}.

\item {\bf Legal and Contractual Challenges}. Legal ambiguities, particularly in international or multi-party agreements, are an open issue to be solved ensuring that SLAs are legally enforceable and aligning them with local regulations \cite{wagle2016cloud}.

\end{itemize}

\section{Summary and Conclusion}
\label{sec:conclusion}

An SLA is a formal document in which a provider defines its level of quality of service assurance through the parameters of the non-functional requirements, usually related to availability and performance. Recently, especially in the context of Cloud computing and the Internet of Things, researchers have investigated new approaches involving the integration of security metrics in SLA to guarantee security assurance and transparency.

\begin{figure}[ht]
	\centerline{
        \includegraphics[scale=0.8]{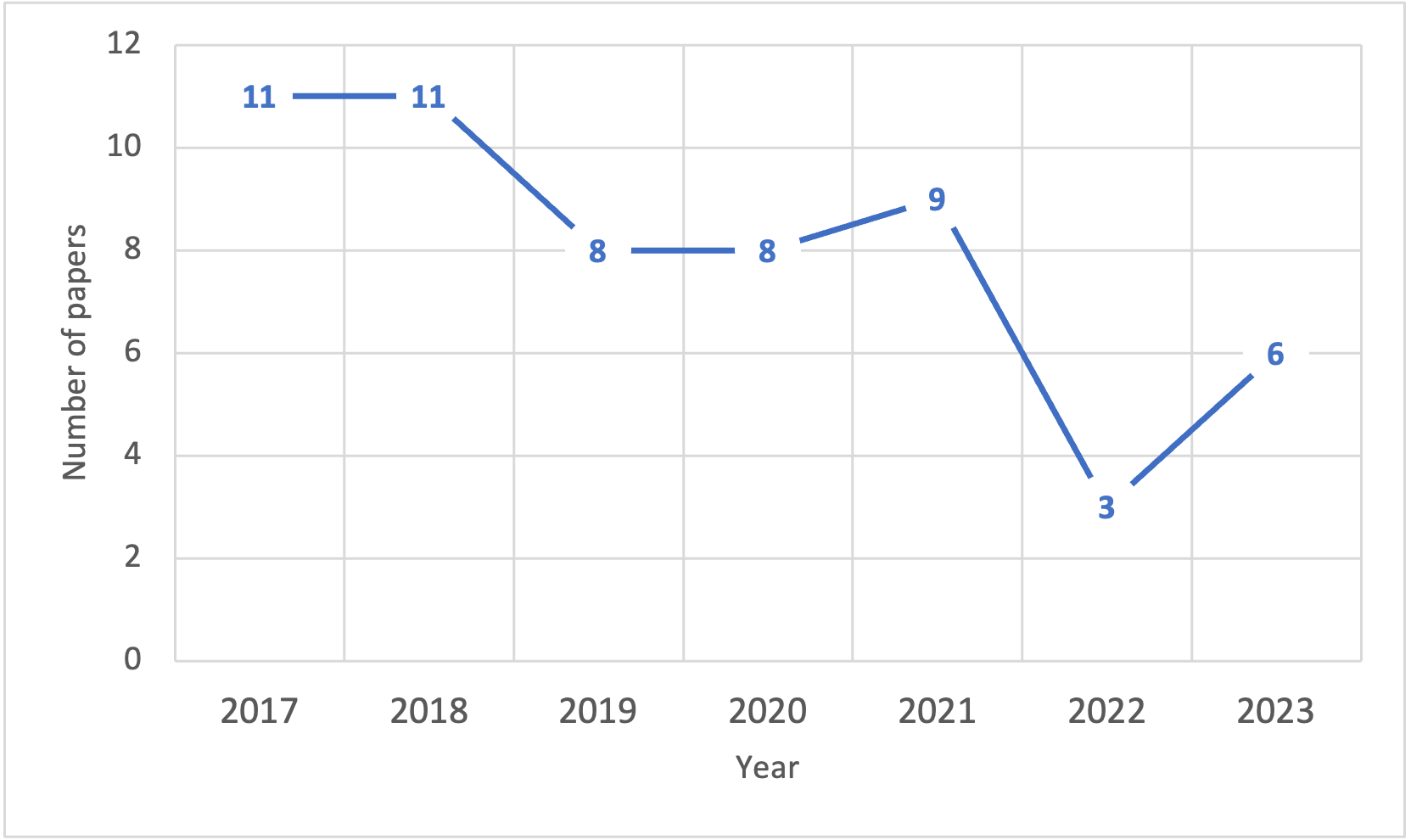}
    }
    \caption{Literature timeline} \label{fig:summaryPerYear}
\end{figure}

\begin{figure}[ht]
	\centerline{
        \includegraphics[scale=0.8]{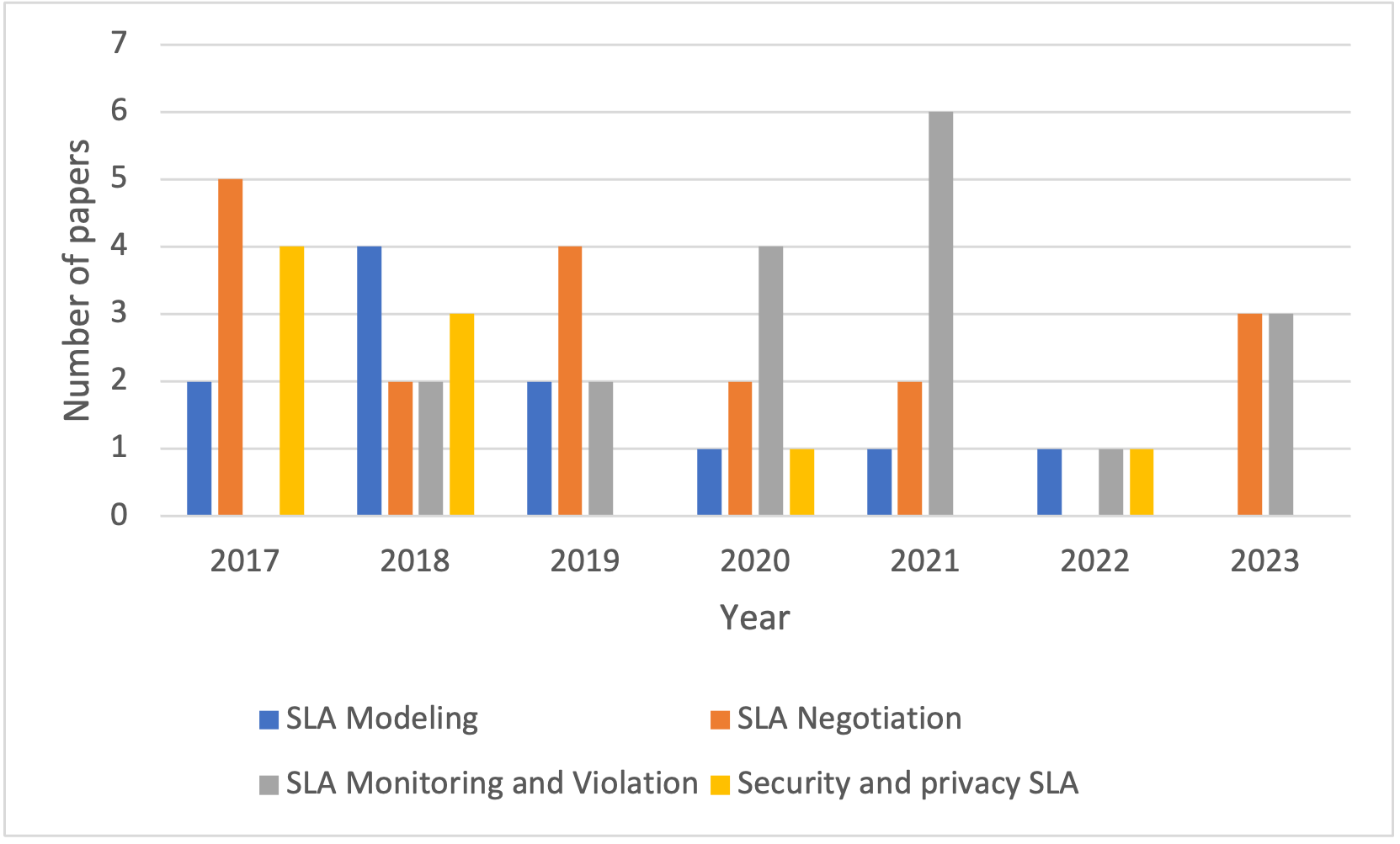}
    }
    \caption{Literature timeline per topic}\label{fig:summaryPerTopicYear}
\end{figure}

To highlight a contribution in this setting, in this survey paper, we provided a detailed review of the most meaningful research papers focusing on SLA management and Security SLA. Following our proposed selection criteria,
we comprehensively examined the most relevant and recent works, including those dealing with the definitions and principles of an SLA, language specifications, and the main steps of its life cycle. Furthermore, this survey paper also addresses the topics of security and privacy SLA management. 
Finally, we discussed the main challenges and future research directions in this context.
In summary, we analyzed $56$ research articles published in renowned international conferences, journals, symposiums, and workshops with a focus on SLA and SecSLA management and related areas. Table \ref{tab:summaryPerTopic} and Figure \ref{fig:summaryPerTopicYear} illustrate a quantitative overview of the reviewed literature divided into topics, whereas Figure \ref{fig:summaryPerYear} visualizes the total analyzed number of articles published per year.

As visible in both the table and two figures above, we can affirm that the analyzed papers are
distributed according to our classification criteria, as follows: SLA modeling ($20\%$), SLA negotiation($32\%$), SLA monitoring and violation ($32\%$), security and privacy SLA ($16\%$). The
limited attention given to the latter phase, whose percentage has decreased to $6\%$ in the last three years, may be attributed to the exploratory nature of most research efforts.

\begin{table}
\centering
\caption{Amount of papers analyzed per topic}
\begin{tabular}{|l|l|}
\hline
    \textbf{Topic} & \textbf{Amount of papers} \\
    \hline \hline
    SLA Modeling & 11\\
    SLA Negotiation & 18\\
    SLA Monitoring and Violation & 18\\
    Security and Privacy SLA & 9\\
    \hline\hline

\end{tabular}
\label{tab:summaryPerTopic}
\end{table}

We plan to continue our study by diving into particular aspects only mentioned in this survey. For instance, an interesting direction can be the review of the papers exploiting existing SLA industrial platforms to give the reader a more extensive spectrum of various issues about them. Moreover, an exhaustive description of all the application domains in the context of IoT can be also a challenging task. 

We hope that this survey can assist both academic researchers and industrial practitioners in apprehending the key features of this domain, highlighting the most significant advancements and perspectives, and encouraging them to undertake this direction in future research advancements.

\section*{Acknowledgments}
This work was supported in part by the project SERICS (PE00000014) under the NRRP MUR program funded by the EU-NGEU, and by the PRIN Project ``HOMEY: a Human-centric IoE-based Framework for Supporting the Transition Towards Industry 5.0'' (code 2022NX7WKE) funded by the European Union - Next Generation EU.

%Bibliography

\end{document}